\documentclass[twocolumn,txfonts,showpacs,nofootinbib,superscriptaddress,preprintnumbers,amsmath,amssymb,epsfig,color]{aa}

\usepackage{natbib}
\bibpunct{(}{)}{;}{a}{}{,} % to follow the A&A style...

\usepackage[usenames]{color}
\usepackage{graphicx}% Include figure files

\usepackage{units, subfigure,amssymb,amsmath}
\usepackage[latin1]{inputenc}

\begin{document}
\input epsf
\title{Disentangling cosmic-ray and dark-matter induced $\gamma$-rays in galaxy clusters}
\titlerunning{$\gamma$-rays in galaxy clusters from a stacking analysis}
\author{
       D. Maurin\inst{1}
	\and C. Combet\inst{1}
  \and E. Nezri\inst{2}
  \and E. Pointecouteau\inst{3}
} 

\offprints{D. Maurin, {\tt dmaurin@lpsc.in2p3.fr}}

\institute{
  Laboratoire de Physique Subatomique et de Cosmologie,
      Universit\'e Joseph Fourier Grenoble 1/CNRS/IN2P3/INPG,
      53 avenue des Martyrs, 38026 Grenoble, France
  \and
  Laboratoire d'Astrophysique de Marseille - LAM, Universit\'e d'Aix-Marseille
\& CNRS, UMR7326, 38 rue F. Joliot-Curie, 13388 Marseille Cedex 13, France
   \and
   Universit\'e de Toulouse (UPS-OMP), Institut de Recherche en Astrophysique et Planetologie,
   CNRS, UMR 5277, 9 Av. colonel Roche, BP 44346, F 31028 Toulouse cedex 4, France
}

\date{Received / Accepted}% It is always \today, today,

\abstract
%Context
{
Galaxy clusters are among the best targets for indirect dark matter
detection in $\gamma$-rays, despite the large astrophysical background
expected from these objects. Detection is now within reach of current
observatories (Fermi-LAT or Cerenkov telescopes); however, assessing the
origin of this signal might be difficult.
}
%Text of aims
{
We investigate whether the behaviour of the number of
objects per `flux' bin ($\log N-\log F$) and that of
the stacked signal could be used as a signature of the
dominant process at stake.
}
%Text of methods
{
We use the \textsc{Clumpy} code to integrate the signal from
decaying or annihilating dark matter and from cosmic rays along the line of
sight. We assume the standard Navarro-Frenk-White (NFW) profile for the dark matter density
and rely on a parametrised emissivity for the cosmic-ray component. In
this context, the consequences of stacking are explored using the MCXC
meta-catalogue of galaxy clusters.
}
%Text of results
{
We find the value of the slope of the $\log N-\log F$ power law
(or the increase of the signal with the number of stacked objects)
to be a clear diagnosis to disentangle decaying
dark matter from cosmic-ray induced $\gamma$-rays. For dark
matter annihilation, depending on the signal boost from the
substructures, it is either similar to the cosmic-ray (CR) signal
(no boost) or similar to the decay case (large boosts). The shift between the
brightest object and its followers also depends on the signal origin.
For annihilation, this shift and the stacked signal are
poorly constrained because of the large uncertainty affecting the boost.
We also underline that the angular dependence of the annihilation
signal is not universal because of the substructure contribution.
}
{}


\maketitle

\section{Introduction}

Contamination by astrophysical backgrounds is the curse of dark matter
(DM) indirect detection searches. Ways around this issue are
multi-wavelength  \citep{2008PhRvD..78d3505R} or/and multi-messenger 
\citep{2009PhRvD..80j3510P} analyses. In $\gamma$-rays, background-free
objects such as dwarf spheroidal galaxies (dSphs) are another option
\citep{2011ApJ...733L..46W}.  For targets presenting an astrophysical background,
a useful diagnosis comes from the spectral behaviour, which differs for
DM-induced or CR-induced $\gamma$-rays: DM signals should exhibit a sharp
cut-off at the DM mass  \citep{2011PhRvD..84j3525B}, whereas
astrophysical processes produce power laws with a slope $\sim 2-3$ (the
cut-off energy depends on the source). For instance, the Galactic centre
signal observed by the High Energy Stereoscopic System H.E.S.S. 
\citep{2004A&A...425L..13A} is consistent
with astrophysical sources. Despite their astrophysical background,
galaxy clusters are also promising DM targets in $\gamma$-rays (or at radio
and microwave frequencies) as first proposed in
\citet{2006A&A...455...21C}. Cerenkov instruments have set constraints
on a DM signal or astrophysical background from the non-detection in
several of these objects: Perseus and Abell 2029 from Whipple
\citep{2006ApJ...644..148P}; Abell 3667 and 4038 from Kangaroo
\citep{2009ApJ...704..240K}; Coma, Abell 496 and 85 from H.E.S.S.
\citep{2009A&A...495...27A,2009A&A...502..437A}. A broader survey has
been made by the Fermi-LAT instrument, where 33 (resp. 49) clusters were
analysed in \citet{2010ApJ...717L..71A} (resp.
\citealt{2012JCAP...07..017A}). The consequences for
hadronic models are discussed, e.g. in \citet{2011ApJ...728...53J}.
Recently, a detection was reported by the MAGIC collaboration in the
Perseus Cluster. However, this signal, also detected by Fermi-LAT
\citep{2009ApJ...699...31A}, is consistent with an emission from the
giant elliptical galaxy NGC 1275 lying at its centre
\citep{2010ApJ...710..634A,2012A&A...539L...2A}. From the Fermi-LAT
results, \citet{2010A&A...519A..82C} argue that it should be  possible
to resolve and detect the diffuse $\gamma$-ray flux of astrophysical
origin coming from the outer corona of the Perseus cluster. Even more
recently, \citet{2012arXiv1201.1003H} reported a detection for Virgo,
Fornax, and Coma from an analysis of Fermi-LAT data. These authors
find that an interpretation in terms of DM annihilation is preferred by
their analysis, although a CR origin is also possible.

From these promising results, we anticipate that many clusters will be
detected with the current or the next generation of $\gamma$-ray
instruments. Multi-wavelength analyses, search for a spectral feature,
and angular dependence of the signal are tests to disentangle
$\gamma$-rays of astrophysical or DM origin. In this paper, we
investigate another test based on the study of a population of sources,
thus somehow positioned between single source and full sky
power-spectrum analyses \citep{2006PhRvD..73b3521A}.  Actually,
several studies have compared the astrophysical and DM annihilation
signal \citep{2009PhRvD..80b3005J,2011PhRvD..84l3509P} as well as DM
annihilation and decay \citep{2012JCAP...01..042H} on a somewhat
limited number of clusters (due to the size of the available catalogues at
the time). However, the systematic study of the prospects of a stacking
analysis with respect to single-source analyses has not been
investigated before. 

This paper is part of an ongoing effort to address this question in
the context of the recently assembled MCXC meta-catalogue of 1743 X-ray
clusters \citep{piffaretti11}. A first paper in this series highlighted
the potential  improvement (a factor $\gtrsim 5-100$) that can be
brought by a stacking analysis over a single-source analysis for the DM
decay case \citep{2012PhRvD..85f3517C}. A second paper considered the
DM annihilation case in the light of the sensitivity of existing and
future $\gamma$-ray instruments \citep{2012MNRAS.425..477N}. It
showed that stacking only brings a factor of $\sim2$ improvement in
sensitivity for the Fermi-LAT instrument, but no improvement for
Cherenkov Telescope Array (CTA)-like observatories. This third and
last paper of the series inspects the case of the CR signal to address the possibility of using
a stacking strategy to disentangle it from DM-induced signals. For that
purpose, we use a generic description of the CR signal and refer the
reader to the two previous papers
\citep{2012PhRvD..85f3517C,2012MNRAS.425..477N} for a detailed
description of the DM modelling. We stress that this study 
remains at the phenomenological level. We are waiting either for the
detection of several objects or a better knowledge of the input
ingredients in order to quantify further the potential
of the new diagnosis we investigate.

We present the ingredients and the calculation in
Sect.~\ref{sec:ingredients}. We then present the results in
Sect.~\ref{sec:results} and conclude in Sect.~\ref{sec:conclusions}.

\section{Ingredients and calculation}
\label{sec:ingredients}

The total flux expected in a given direction $(l,b)$ (Galactic
coordinates) and integrated over the solid angle $\Delta\Omega$ is
given by the product of an energy-dependent term with an astrophysical
term ${\cal A}$,\footnote{The redshift distribution of the MCXC
catalogue of galaxy clusters \citep{piffaretti11} peaks at $z\sim 0.1$
(see their Fig.~1). Following \citet{2012PhRvD..85f3517C} and
\citet{2012MNRAS.425..477N}, we neglect the absorption for the MCXC
galaxy clusters, so that the energy-dependent term and the spatial term are
decoupled.} 
\begin{equation}
  \frac{d\phi(E,l,b,\Delta\Omega)}{dE} = \frac{dN}{dE}(E) \times {\cal A}(l,b,\Delta\Omega)\,.
\end{equation}
In this study, we discard the spectral term and focus on ${\cal A}$,
which encodes all the information about the spatial dependence of the
signal, and the relative intensity between clusters. ${\cal A}$ will be
termed `flux' in the following, but should be understood as the
astrophysical contribution to the actual $\gamma-$ray flux. For an
observation in the direction of the cluster's centre $(l_{\rm
cl},b_{\rm cl})$ and using an integration angle $\alpha_{\rm int}$,
\begin{equation}
\label{eq:Aterm}
{\cal A}_X(\alpha_{\rm int})\equiv{\cal A}_X(l_{\rm cl},b_{\rm cl},\Delta\Omega)=\int_{\Delta\Omega}\int {\cal E}_X(l',\Omega) \,dl'd\Omega
\end{equation}
is the integral of the `emissivity' ${\cal E}_X(l',\Omega)$, over line
of sight $l'$ and solid angle $\Delta\Omega =
2\pi\cdot(1-\cos(\alpha_{\rm int}))$ for the process $X$. For DM,
${\cal E}_{\rm Decay}=\rho_{\rm DM}$ for decay and ${\cal E}_{\rm
Annihil.}=\rho_{\rm DM}^2$ for annihilation, where $\rho_{\rm DM}$ is
the DM density profile in the galaxy cluster, as presented in
Sect.~\ref{sec:NFW}. For CR, ${\cal E}_{\rm CR}=C_{\rm CR}$, as defined
in Sect.~\ref{sec:CR}.

The integral Eq.~(\ref{eq:Aterm}) is computed for all MCXC clusters with
the \textsc{Clumpy}\footnote{\tt http://lpsc.in2p3.fr/clumpy/} code 
v2011.09 \citep{2012CoPhC.183..656C} adapted to
include the CR-induced $\gamma$-ray case described below.

\subsection{Dark matter halos}
\label{sec:NFW}
The DM distribution for each galaxy cluster is parametrised from
(see \citealt{2012PhRvD..85f3517C} for more details):
\begin{itemize}
 \item its $M_{500}$ value\footnote{$M_\Delta$ (with $\Delta=200$ or
    500) is the mass enclosed in a sphere of radius $R_\Delta$, the
    radius within which the average density reaches $\Delta$ times the
    critical density of the Universe.} provided in the MCXC catalogue
    \citep{piffaretti11};
 \item the choice of a universal DM profile, here the widely used NFW
    profile \citep{navarro97}, 
    \begin{equation} 
    \displaystyle
    \rho_{\rm NFW}(r)=\frac{\rho_s}{\left(\frac{r}{r_s}\right)\left(1+\frac{r}{r_s}\right)^2}\,,
    \label{eq:rhoNFW}
    \end{equation}
    where $r_s$ is the scale radius and $\rho_s$ is the normalisation;
 \item a mass-concentration relationship, where the concentration is
    defined to be $c_\Delta\equiv R_\Delta/r_s$ for an NFW
    profile\footnote{In practice, we work with virial quantities
    (see, e.g. \citealt{2010MNRAS.404..502G}, to switch from $\Delta$
    to virial quantities), e.g. $R_{\rm vir}$ is the physical size of the
    galaxy cluster halo.}. The latter is
    observationally constrained at the cluster scale
    \citep{2005A&A...435....1P,2007ApJ...664..123B,2010A&A...524A..68E},
    and it has also been extensively studied in numerical simulations
    \citep{2002ApJ...568...52W,2006ApJ...652...71W,
    2003ApJ...597L...9Z,2009ApJ...707..354Z,2007MNRAS.381.1450N,2008MNRAS.387..536G,
    2011MNRAS.410.2309G,2008MNRAS.391.1940M,2008MNRAS.390L..64D,2010MNRAS.405.2161D,
    2010MNRAS.404..502G,2011MNRAS.411..584M, 2011ApJ...740..102K}. In
    this study, we use the parametrisation of
    \citet{2008MNRAS.390L..64D}.
   \end{itemize}

For DM annihilation, we also have to take into account the contribution
from substructures as they have been shown to boost the signal. In this
work, we assume the following  for the mass and spatial distribution of
the substructures: i) $dN_{\rm subs}/dM\propto M^{-\alpha_{\rm M}}$ with
$\alpha_{\rm M}=1.9$, a mass fraction $f_{\rm DM}=10\%$ in substructures
\citep{2008MNRAS.391.1685S}, a minimal and maximal mass of
$10^{-6}~M_\odot$ and $10^{-2}M_{\rm cluster}$ respectively, and the
\citet{2001MNRAS.321..559B} concentration (down to the minimal mass);
ii) the substructure spatial distribution $dN_{\rm subs}/dV$  follows
the host halo smooth profile. Recent high-resolution numerical
simulations that are for for Galaxy-like objects \citep{2008MNRAS.391.1685S}
or are dedicated to cluster-size halos \citep{2012MNRAS.425.2169G} provide
different figures for the slope $\alpha_{\rm M}$, the mass fraction
$f_{\rm DM}$, and the spatial distribution of substructures. We will
briefly comment on how other choices impact on our
analysis\footnote{A detailed discussion of the annihilation signal and
its dependence on the substructure parameters can be found in Sect.~3.4
of \citet{2012MNRAS.425..477N} and is not repeated here. We underline
that taking the same distribution as the smooth halo for this quantity
or using that of \citet{2012MNRAS.425.2169G} does not impact the
annihilation results (see \citealt{2012MNRAS.425..477N}).}.

\subsection{Cosmic-ray component}
\label{sec:CR}

High-energy $\gamma$-rays can also be produced from astrophysical
processes, in particular from the interaction of cosmic-ray protons
with the gas of the cluster. \citet{2010MNRAS.409..449P} found a
universal radial dependence of the emissivity ${\cal E}_{\rm CR}\equiv
C_{\rm CR}(r)$ of the clusters based on cosmological simulations. These
authors acknowledge, however, that the CR spatial distribution from which
$\gamma$-rays are emitted could be affected by, e.g. additional CRs
injected from AGN, or CR diffusion in momentum and space. For
instance, \citet{2011A&A...527A..99E} argue that merging clusters
should have a more centrally concentrated CR population than relaxed
ones, providing bi-modality of their $\gamma$-ray emissivities.
Nevertheless, we limit ourselves to the simplest case and assume a
universal dependence across the cluster population we study here.  We
follow the formalism from \citet{2010MNRAS.409..449P} and
\citet{2011PhRvD..84l3509P}:
\begin{equation}
   \label{eq_CRR}
   C_{\rm CR}(r) = \left\{\frac{c_{200}-5\cdot 10^{-7}}{1+(\frac{r}{R_{\rm trans}})^{-b}} +5\cdot 10^{-7} \right\}
      \times \rho_{\rm gas}^2(r)\,,
\end{equation}
\begin{equation}
 c_{200} = 1.7 \cdot 10^{-7} \left(\frac{M_{200}}{10^{15} M_\odot}\right)^{0.51}\!\!\!\!,\,\,\;\;
   b = 1.04 \left(\frac{M_{200}}{10^{15} M_\odot}\right)^{0.15}\,,\nonumber
\end{equation}
\begin{equation}
     {\rm and~} R_{\rm trans} = 0.021 \, R_{200} \left(\frac{M_{200}}{10^{15} M_\odot}\right)^{0.39}\,. \nonumber
\end{equation}

Similar to the X-ray luminosity, the $\gamma$-ray emissivity is
proportional to the square of the gas density $\rho_{\rm gas}^2(r)$. It
is thereby very sensitive to the densest central parts of clusters. The
relation between the DM and $\rho_{\rm gas}(r)$ has been extensively
studied from a theoretical point of view
\citep{1998ApJ...497..555M,1998ApJ...509..544S,2001MNRAS.327.1353K,
2005ApJ...634..964O,2009ApJ...700..989B,2009ApJ...700.1603F,
2012MNRAS.422..686C}.  From the REXCESS representative sample,
\citet{2008A&A...487..431C} showed the universal behaviour of
$\rho_{\rm gas}(r)$ in clusters (though a large scatter exists at the
centre). The AB model, first introduced by \citet{2002A&A...394..375P},
and given by
\begin{equation}
\label{eq:gas}
   \rho_{\rm gas}(x) \propto \left(\frac{x}{x_c}\right)^{-k_1} 
     \left[ 1+\left(\frac{x}{x_c}\right)^2\right]^{-3k_2/2 + k_1/2}
\end{equation}
with $x\equiv r/R_{500}$, was fitted to the average REXCESS density
profile by \citet{arnaud10}, with best-fit parameters $x_c=0.303$, $k_1=0.525$,
and $k_2=0.768$. This analytical phenomenological model
reproduces well the cuspy shape of density profiles derived from X-ray
observations and was also used for the MCXC assembly
\citep{piffaretti11}. In order to remain consistent with respect to the
data we use, the same AB profile is used here. The normalisation of
Eq.~(\ref{eq:gas}) is chosen to give $f_{\rm gas}M_{500}$ at $R_{500}$.
From the MCXC masses $M_{500}$ and the empirical relation between
$f_{\rm gas}$ and $M_{500}$ given for the REXCESS sample
\citep{2009A&A...498..361P}, an average value of $f_{\rm gas}=0.10\pm
0.06$ is derived, which we use for the whole MCXC sample. We note that the
gas fraction increases with radius in halos (to ultimately tends
towards the cosmic value), and the intrinsic scatter is important and
scales with the cluster mass 
\citep{vikhlinin06,2009A&A...501...61E,2009A&A...498..361P,2010A&A...511A..85P}.

\section{Results}
\label{sec:results}

We first comment on the universality of the  $\gamma$-ray signal
angular dependence. In particular, we inspect whether this universality
holds for any signal hypothesis  (annihilating DM, decaying DM, or
CRs); in that case, observations of different objects may be optimally
stacked to increase the signal. We also inspect whether a given angular dependence
can be associated to a unique signal origin. We then turn to the
comparison of the galaxy cluster population signals.

\subsection{A different angular dependence for different origins?}

The spatial dependence of the signal has been suggested as a test to
disentangle DM decay from annihilation, in particular in dSphs
\citep{2010JCAP...07..023P}. Indeed, the use
of universal DM profiles ensures that the projected profiles are
universal if properly normalised by their value at $r=r_s$. By
construction, the projected profiles for decay are universal, and so
they are for annihilation if the smooth signal only is taken into
account. However, as can be deduced from \citet{2012MNRAS.419.1721G}
results and as underlined in \citet{2012MNRAS.425..477N}, the
substructure contribution overtakes the smooth contribution at various
distances from the centre, depending on the mass of the object rather than
on its scale radius. Small objects (like dSphs) are
dominated by their smooth contribution up to large distances from their
centre, whereas massive galaxy clusters are almost completely dominated
by the substructure contribution. As a result, departure from
universality of the total DM annihilation signal (the one which we have
access to) is expected.

\begin{figure}[!t]
\begin{center}
\includegraphics[width=\columnwidth]{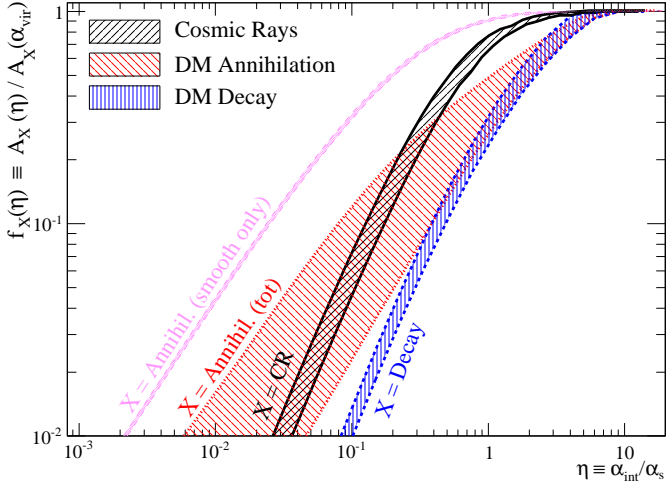}
\caption{Signal fraction $f_X(\eta)$ (for process $X$) as function of
the reduced integration angle $\eta\equiv \alpha_{\rm int}/\alpha_s$.
For the sake of presentation, only the envelopes for the 1743 clusters
of the catalogue are shown. For annihilation, the signal for the
smooth only is also shown (the substructure contribution only is similar
to the decay case).}
\label{fig:fig1}
\vspace{-0.5cm}
\end{center}
\end{figure}

This is illustrated using all MCXC galaxy clusters. Defining for any
cluster its virial angle $\alpha_{\rm vir}=\tan^{-1}(R_{\rm vir}/d)$
and its scale angle $\alpha_s=\tan^{-1}(rs/d)$ ($d$ its distance to the
observer), we show in Fig.~\ref{fig:fig1} the fraction of the
integrated signal with respect to the maximal signal obtained, i.e.
\begin{equation} 
  f_X(\eta) \equiv 
     \frac{{\cal A}_X(\alpha_{\rm int}= 
      \eta\cdot\alpha_s)}{{\cal A}_X(\alpha_{\rm vir})}
\end{equation}
as a function of $\eta=\alpha_{\rm int}/\alpha_s$.

For DM annihilation, the smooth contribution (left-hand envelope) is the
most centrally concentrated as it comes from the integration of the
steep DM profile squared ($\rho^2_{\rm DM} \stackrel{r\rightarrow
0}{\longrightarrow} r^{-2}$). 

For the decay signal (right-hand envelope), the external parts of the
halo contribute more significantly to the signal because the quantity to
integrate is shallower ($\rho_{\rm DM} \stackrel{r\rightarrow
0}{\longrightarrow}  r^{-1}$). We note that in the MCXC catalogue, the mass
ratio of the most massive galaxy cluster to the least massive one is
$\sim 100$, so that these clusters have different concentrations
$R_{\rm vir}/r_s$, hence a different rescaled angular size $\alpha_{\rm
vir}/\alpha_s$. It does not impact on the DM annihilation signal from
the smooth halo (because this signal is strongly centrally
concentrated), but it leads to a small spread for a decay-like
signal. We checked that using an Einasto profile
\citep{2008MNRAS.391.1685S} instead of a NFW for the DM halos does not
change the curves significantly. 

For CRs (black envelope), the trend is similar to the decay case because
$\rho^2_{\rm gas} \stackrel{r\rightarrow 0}{\longrightarrow} 
r^{-1.05}$. The shift between the CRs and decay envelopes comes from
the gas being more centrally concentrated than DM. 

The DM annihilation from substructures is directly proportional to the
integration of the substructure spatial distribution $dN_{\rm
subs}/dV$, which we have chosen in this work to follow the smooth DM
profile. Because of this choice, the signal from substructures
follows the same dependence as DM
decay and is therefore not repeated in the figure. Similar to the
smooth annihilation signal, changing this distribution does not impact on
the conclusions significantly. As underlined, the more complex (i.e.
non-universal) scaling of the total annihilation signal shown in
Fig.~\ref{fig:fig1} comes from the fact that both the smooth halo and
the substructures contribute to the signal. As shown in Fig.~4 of
\citet{2012MNRAS.425..477N}, the distribution of boosts for the MCXC
sample depends on the substructure configuration, and it can vary from
1 (no boost) to about 100 for a Phoenix-like configuration. In the
former case, the angular dependence would be close to the smooth only
annihilation case, whereas for the latter case, it would show a
decay-like angular dependence. 

To conclude, whereas the angular dependence is sometimes argued to be a
good diagnosis to disentangle a CR signal from a DM signal, we draw
attention to the fact
that the situation is not that clear cut for galaxy clusters. Indeed, a
CR signal is more centrally concentrated than a DM decay signal.
However, for the annihilation case, it could be more concentrated
than the CR signal, close to the DM decay case, or in-between.

\subsection{Number of objects per flux bin}
\begin{figure}[!t]
\begin{center}
\includegraphics[width=0.49\columnwidth]{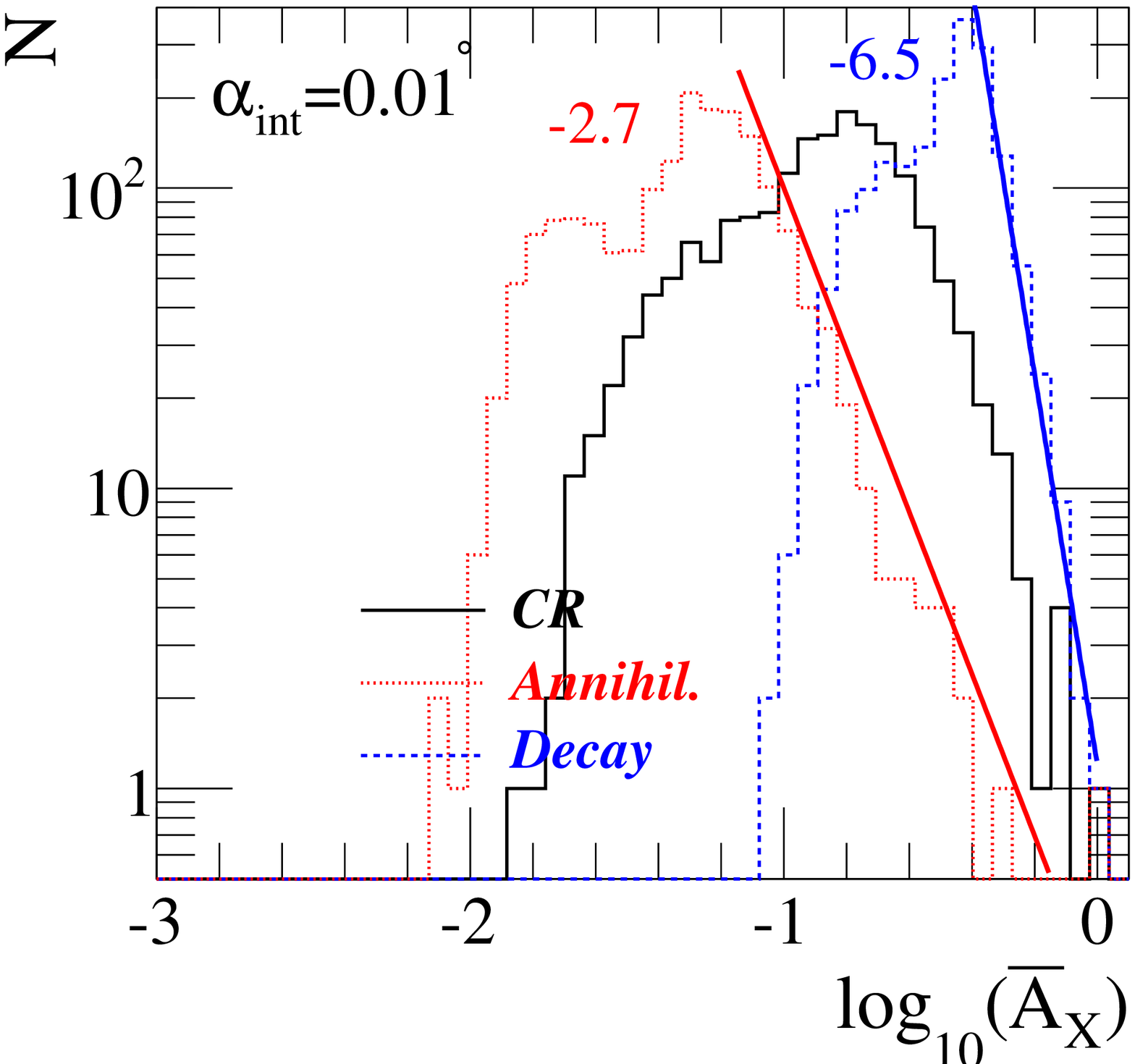}
\includegraphics[width=0.49\columnwidth]{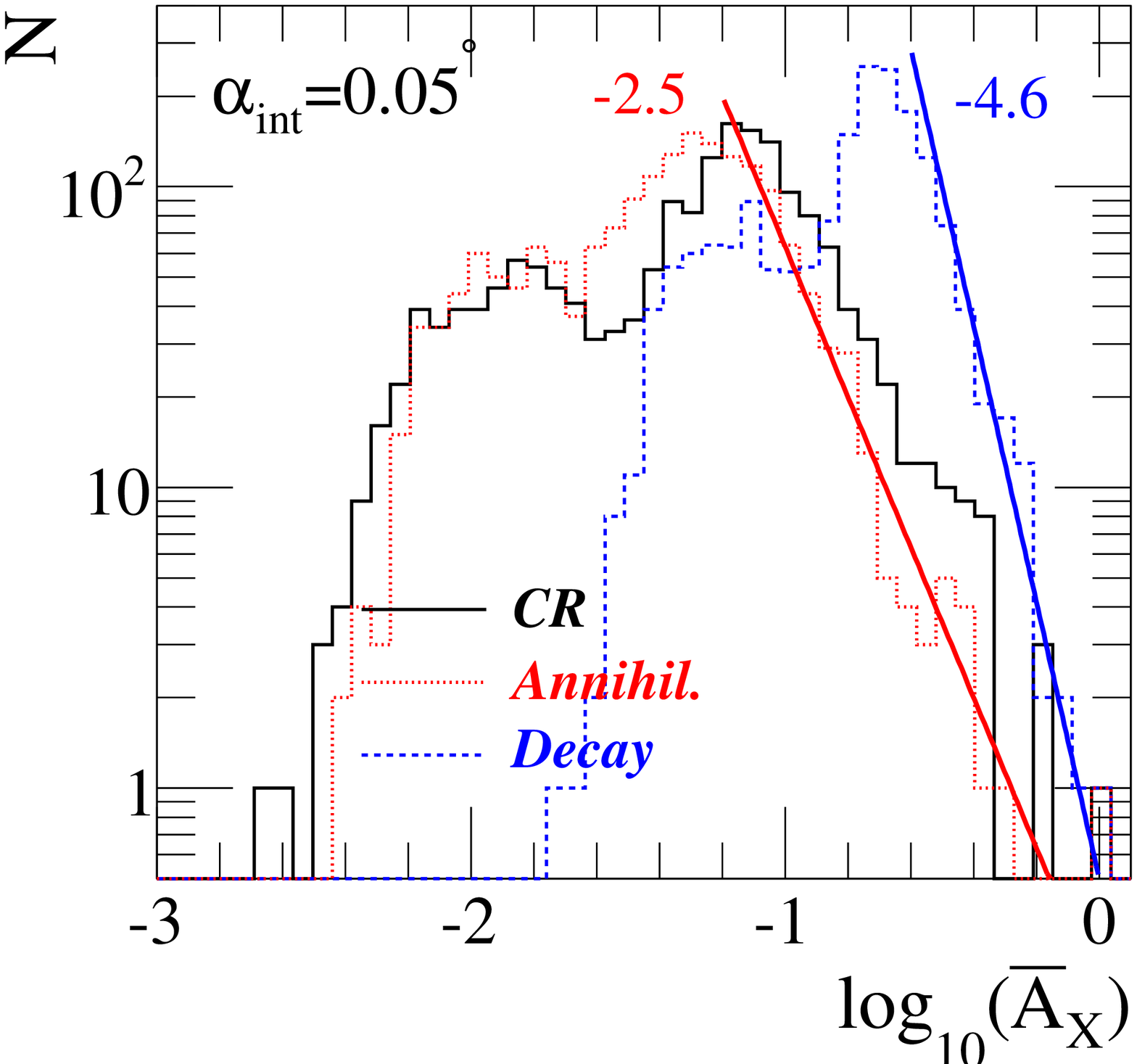}
\includegraphics[width=0.49\columnwidth]{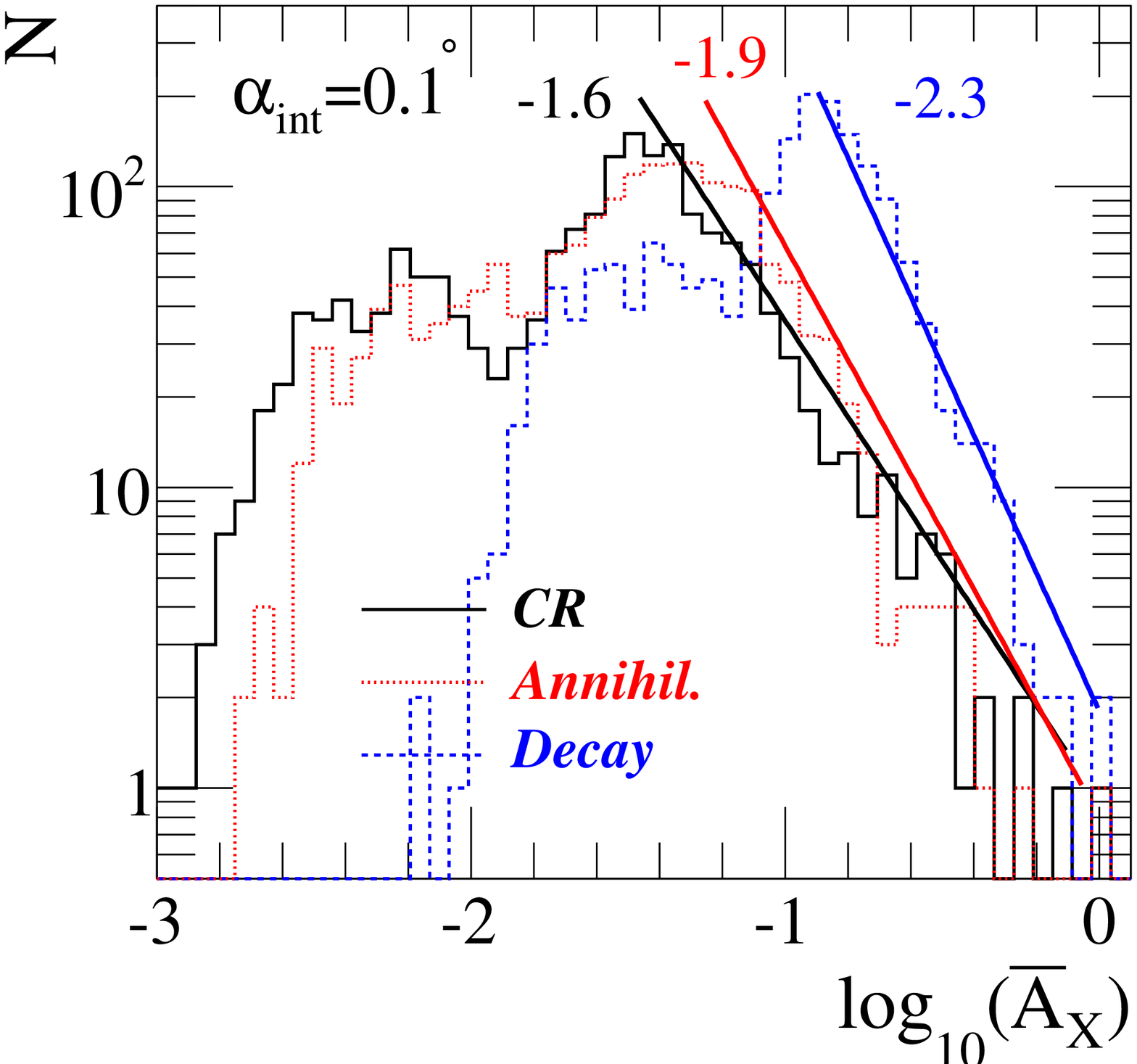}
\includegraphics[width=0.49\columnwidth]{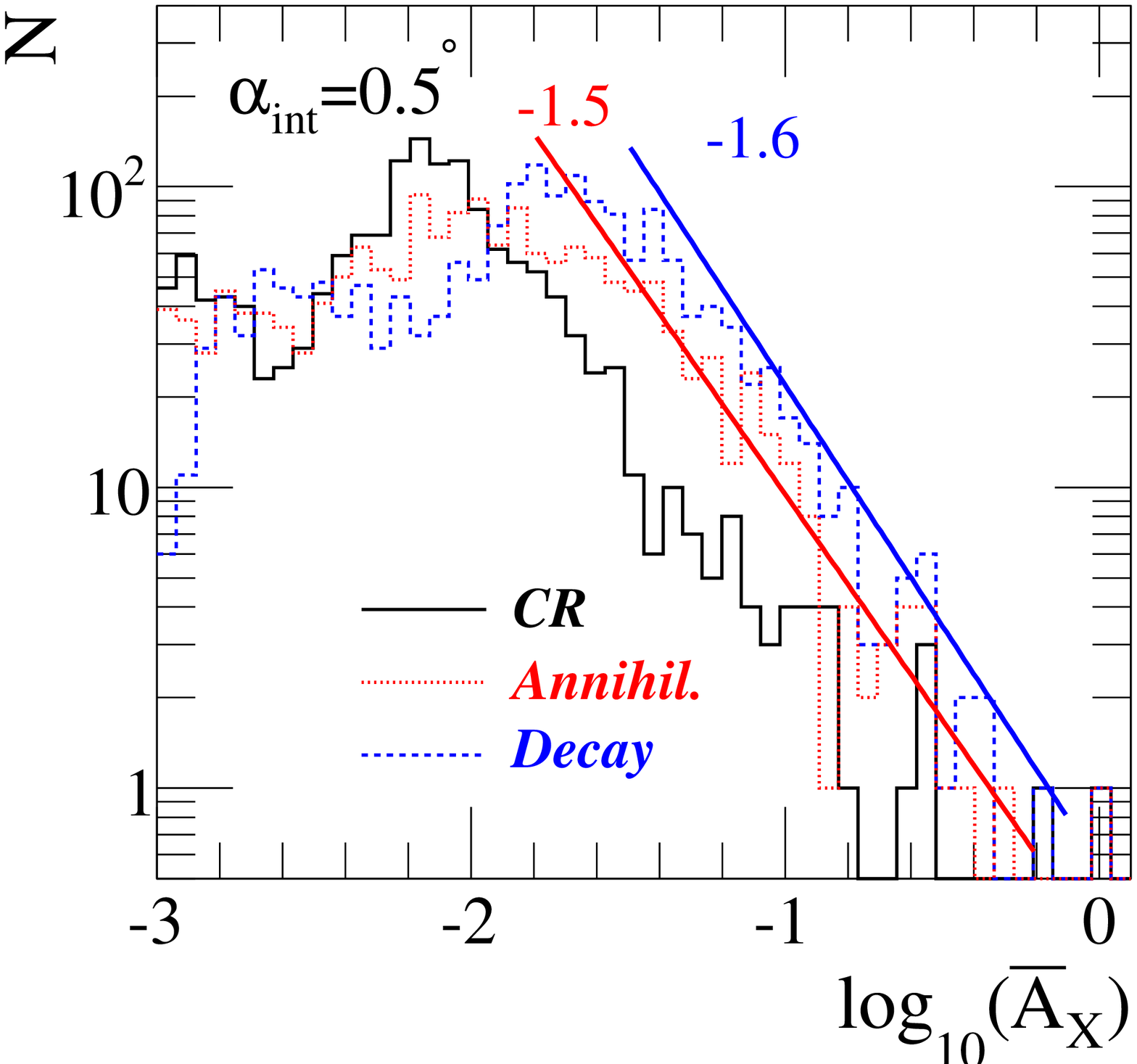}
\caption{Number of galaxy clusters per logarithmic bin of $\bar{A}_X$
(see Eq.~(\ref{eq:lum_rel})), for the CR component (solid black), DM
annihilation (dotted red), and DM decay (dashed blue). The four panels
correspond to four integration angles $\alpha_{\rm int}$ (from top left
to bottom right, $0.01^\circ$, $0.05^\circ$, $0.1^\circ$, and
$0.5^\circ$). The index $s$ of the power law $N\propto {\cal A}^{s}$
fit is shown for $\alpha_{\rm int}=0.1^\circ$ for the three
components.}
\label{fig:fig2}
\vspace{-0.5cm}
\end{center}
\end{figure}

We now turn to the behaviour of the number of objects found given
their flux, i.e. the $\log N - \log F$ dependence, where $F \propto {\cal
A}_X$ is the actual observed flux (it is referred to as the
$\log_N-\log{\cal A}_X$ dependence below). Let us define the relative
flux $\bar{A}_X^i$ of the $i-th$ galaxy cluster, i.e. the flux
relative to that of the brightest object of the catalogue for the process
$X$ at stake:
\begin{equation}
  \bar{A}_X^i \equiv \frac{{\cal A}_X^i}{{\cal A}_X^{\rm brightest}}\,.
  \label{eq:lum_rel}
\end{equation}
Such a definition allows us to visually compare the behaviour for the
different $X$ processes. 
\begin{figure*}[!t] 
\begin{center}
\includegraphics[width=\columnwidth]{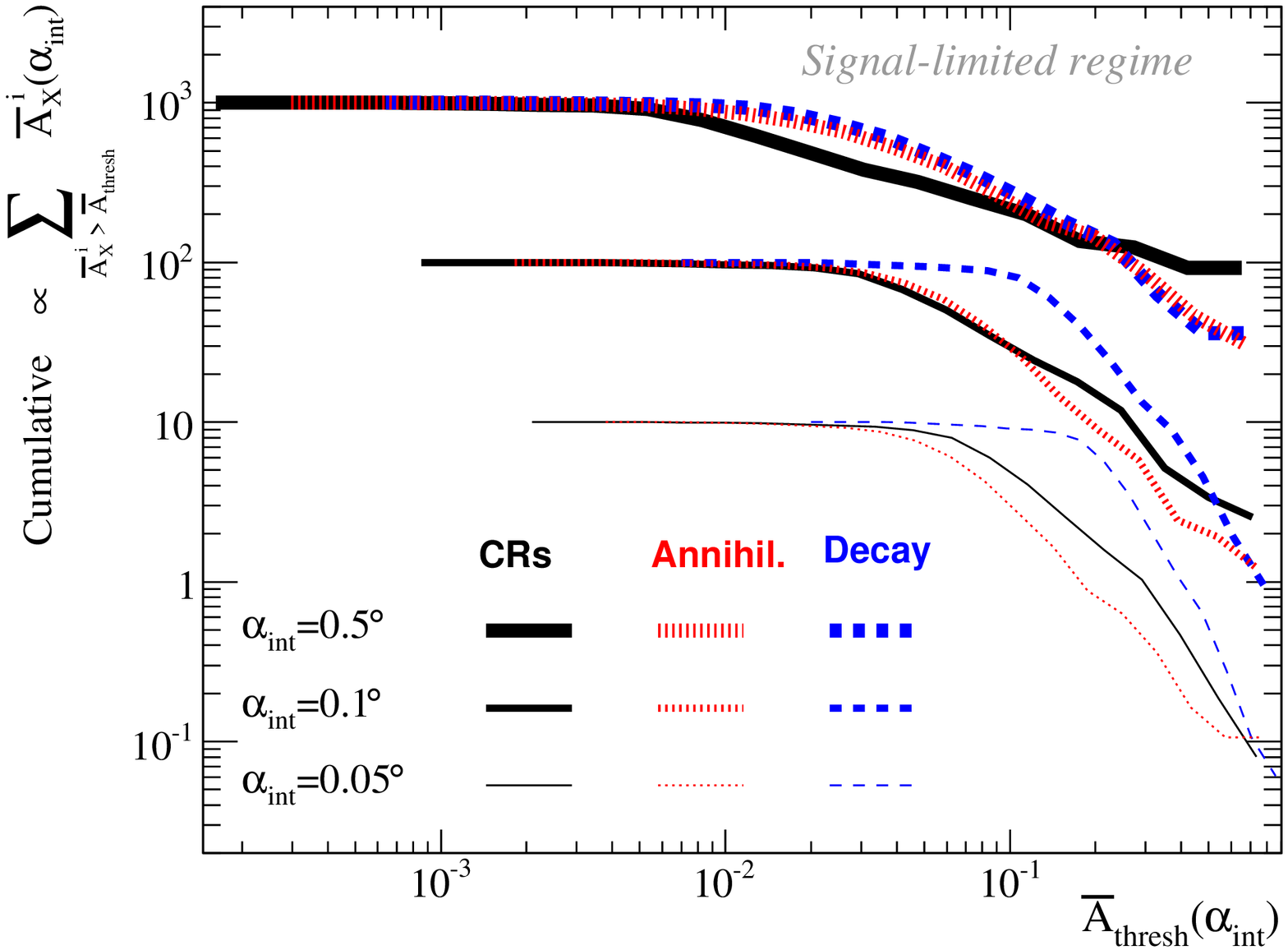}
\includegraphics[width=\columnwidth]{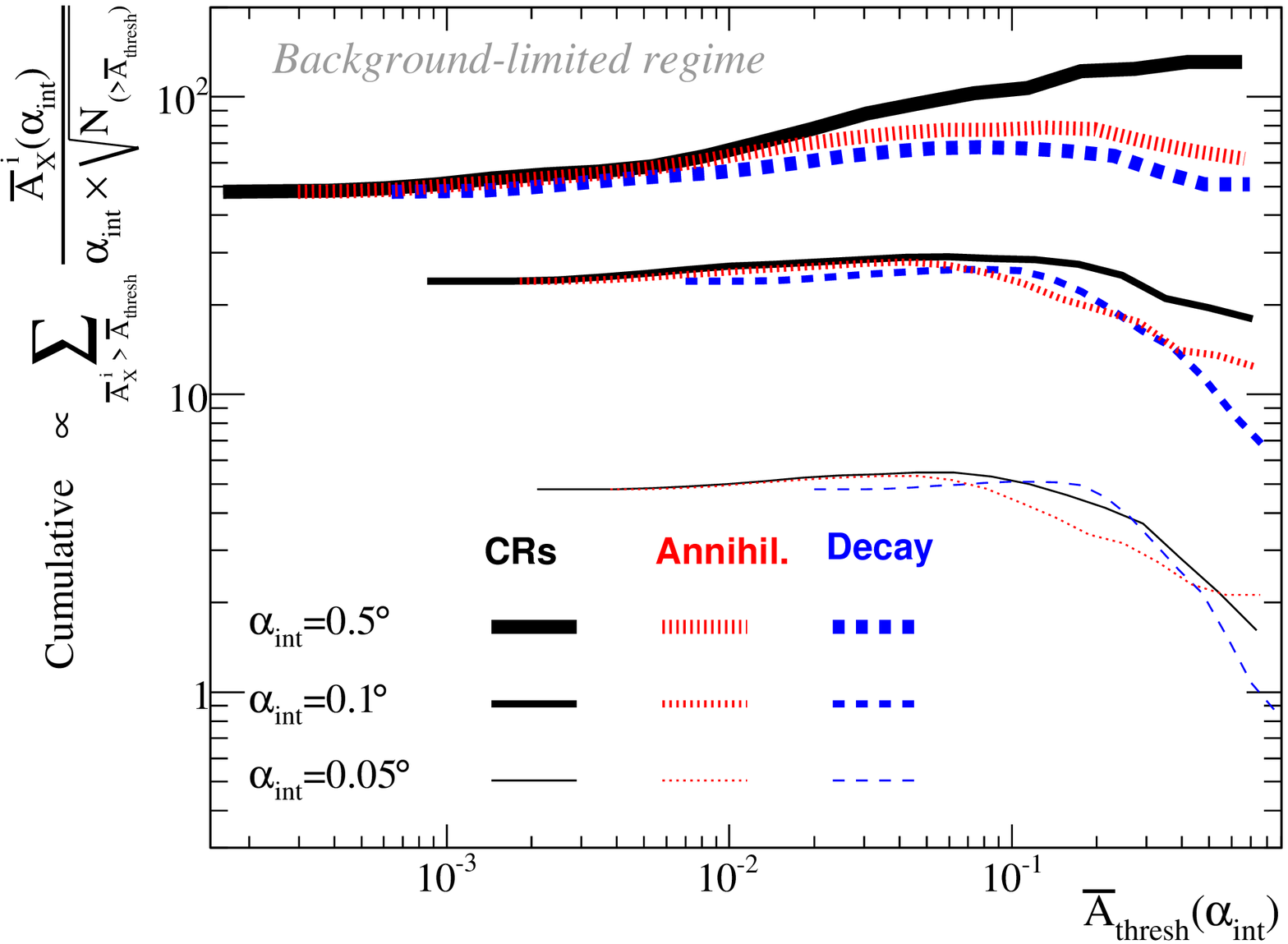} 
\caption{Cumulative of the signal above a threshold value $\bar{A}_{\rm
thresh}$ for the signal-limited regime (left panel, corresponding to
Eq.~(\ref{eq:K1})) and for the background-limited regime (right panel,
corresponding to Eq.~(\ref{eq:K2})). The three line widths correspond to
three integration angles, and the three styles and colours correspond to
CRs, DM annihilation, and decay. An arbitrary normalisation is set in
order to ease the comparison by eye of the different signals.}
\label{fig:fig3}
\vspace{-0.5cm}
\end{center}
\end{figure*}

This is shown in Fig.~\ref{fig:fig2}, where the number of objects $N$
per logarithmic bin (of $\bar{A}_X$) is plotted for the MCXC sample. At
first glance, the value of the slope $s$
($N\propto\bar{A}_X^{-s}\propto{\cal A}_X^{-s}$) in the region beyond
the distribution peak can be used as a diagnosis of the signal
origin\footnote{The catalogue is not complete across the range of masses
it samples, explaining the peaks seen in Fig.~\ref{fig:fig2} (the
selection function is undetermined, beyond the scope of this paper,
and not accounted  for in the discussion).}. We observe that the
behaviour of the DM annihilation and CR components is very similar.
This is tentatively explained by the fact that the CR emissivity Eq.~(\ref{eq_CRR}) is
$\propto \rho_{\rm gas}^2\propto M^2$ and DM annihilation is $\propto
\rho_{\rm DM}^2$. For the latter, the slope $s_{\rm annihil.}$  varies
in the range $[s_{\rm decay}/2,\; s_{\rm decay}]$: for small integration
angles (top left), ${\cal E}_{\rm decay}\propto \rho_{\rm DM}$ and
${\cal E}_{\rm annihil}\propto \rho_{\rm DM}^2$ and $s_{\rm
annihil.}\approx s_{\rm decay}/2$ (substructures scarcely play a role in the
central regions of the cluster), whereas for integration angles
encompassing the whole cluster (bottom right), the substructure
contribution becomes dominant and $s_{\rm annihil.}\approx s_{\rm
decay}$. Actually, the annihilation signal at any given integration
angle is sensitive to the DM substructure configuration (as explained in
the previous section, see also Sect.~3.4 of
\citealt{2012MNRAS.425..477N}), so that the values of $s_{\rm annihil.}$
shown in Fig.~\ref{fig:fig2} are very uncertain (they are bracketed by
the value obtained for the CR case and the value for the decay case).

The actual value of $s$ is difficult to explain from simple
arguments because it comes from the interplay of the profile, normalisation,
and integration angle for each cluster. This analysis nevertheless 
shows that a CR origin can be disentangled from a DM origin in
the case of a decaying DM candidate.

\subsection{Stacked signal}

As deduced from Fig.~\ref{fig:fig3}, the ratio between the
brightest and next-to-brightest clusters  could also be an indicator 
of the $\gamma$-ray signal origin (it is the largest for CR origin and
the lowest for decay at $0.5^\circ$, bottom right-hand panel). This is,
however, sensitive to the individual modelling of the clusters, especially
for the annihilation case because of the extra uncertainty from DM
substructures. In that respect, this ratio criterion (or also using the
ratio of the CR signal to the DM annihilation signal, see
\citealt{2011PhRvD..84l3509P}) should be used with caution. The
cumulative\footnote{The cumulative is defined to be the
cumulated flux of all objects brighter than a threshold flux} 
of the signal also
suffers from similar uncertainties. However, as the number of stacked
objects increases, the sensitivity to the modelling of a given halo
becomes less crucial. The possibility to identify the origin $X$ of a
signal from the behaviour of the stacked signal is very similar in
spirit to comparing the slope $s$ of the $\log N-\log {\cal A}_X$
relation. However, if the signal is only seen by means of stacking, it
may not be possible to get $s$, but still possible to say something
about the origin from this detection.

The promise of a stacking analysis for a signal $X$ depends on the slope
$s$ of the $\log N-\log {\cal A}_X$ relation. If the number of objects
increases faster than the signal decreases, a better sensitivity should
be reached compared to the single-source analysis case. This was shown
to be the case for decay in \citet{2012PhRvD..85f3517C}, but not for
annihilation \citep{2012MNRAS.425..477N}. A detailed discussion of
detectability in the light of the Fermi-LAT and CTA-like instruments was
also presented in that paper. Here, we extend the analysis to the CR
case, and comment on the differences between the various signals. The
simple approach used in \citet{2012PhRvD..85f3517C} is enough to capture
the trends and draw some conclusions, i.e. we consider the generic
extreme case of a signal- and a background-limited
observation\footnote{The detectability of galaxy clusters depends both on the
integrated signal and on the much higher level of charged particle
and diffuse astrophysical $\gamma$-ray backgrounds. In the
background-limited regime the best approach is to maximise the
signal-to-noise ratio. For a uniform background, it corresponds to the
cumulative of ${\cal A}_X$ divided by the square root of the background
which is $\propto \alpha_{\rm int}^2 N$.}. 

As in the previous section, we work with the normalised flux
$\bar{A}_X$ defined by Eq.~(\ref{eq:lum_rel}) to ease the comparison
between the different $X$ origins. The cumulative  for a signal-limited
observation (left-hand panel) corresponds to
\begin{equation}
   {\cal K}_{\rm Sig-lim}(\alpha_{\rm int}) \propto 
    \sum_{\bar{A}_X^i >\bar{A}_{\rm thresh}} \bar{A}_X^i(\alpha_{\rm int}),
\label{eq:K1}
\end{equation}
while that for a background-limited observation is
\begin{equation}
   {\cal K}_{\rm Bkgd-lim}(\alpha_{\rm int})
    \propto \frac{{\cal K}_{\rm Sig-lim}(\alpha_{\rm int})}{\alpha_{\rm int} \sqrt{N_{(>\bar{A}_{\rm thresh})}}}\,,
\label{eq:K2}
\end{equation}
where $N_{(>\bar{A}_{\rm thresh})}$ is the number of clusters
satisfying $\bar{A}_X^i (\equiv {\cal A}_X^i/{\cal A}_X^{\rm
brightest})  >\bar{A}_{\rm thresh}$.  This is shown in
Fig.~\ref{fig:fig3} for the MCXC catalogue\footnote{The plateau reached
by the cumulative signal on the left-hand panel is due to the drop of the
number of objects seen in Fig.~\ref{fig:fig2}.}. For display purposes,
the curves for various processes $X$ are normalised to the asymptotic
value (when all objects are stacked), whereas the curves for different
integration angles $\alpha_{\rm int}$ are shifted from one another.

Regardless of the integration angle and whether the regime is
signal or background limited, stacking leads to a significantly larger
increase of the signal for DM decay than for the CR case. Indeed, as
shown in \citet{2012PhRvD..85f3517C} for the DM decay component,
improvement with respect to the brightest source can reach at best
(with the MCXC catalogue) a factor of $\sim 100$ for background-free
instruments, but a factor $\sim 5$ is expected for more realistic
background-limited instruments. For both the CR and DM annihilation
component, the improvement is respectively $\sim 50$ (background free)
and $\sim 2$ (background limited) for $\alpha_{\rm int}\lesssim
0.1^\circ$. A detailed analysis based on a realistic instrumental
response for the Fermi-LAT and CTA instruments shows that this factor
of two (for annihilation) is only achieved for Fermi-LAT, there
is no improvement with CTA-like instruments
\citep{2012MNRAS.425..477N}. Hence a similar result is expected for the
CR-induced origin. Although in principle the behaviour of
stacked signals at different angles for the DM annihilation case can be
disentangled from the two other cases for the signal-limited regime,
the uncertainties on the substructure distribution are such that the
spread generated on the stacked annihilation signal is encompassed by
the CR and DM decay case and thus cannot be distinguished for now.

\section{Conclusion}
\label{sec:conclusions}

The first two papers of this series investigated the potential
benefit of a stacking analysis of clusters of galaxies for DM detection.
Relying on the largest uniformed X-ray catalogue to date, the MCXC catalogue
\citep{piffaretti11}, it was shown that a DM decay signal
benefits from a stacking analysis \citep{2012PhRvD..85f3517C}, but
that the situation is less promising for the annihilation signal
\citep{2012MNRAS.425..477N}. In this third and last paper of the series,
we extended the analysis to the expected
CR-induced component, hoping to put to the forth a difference that may
be used as a novel diagnosis to identify the origin of the signal when
detection in such objects becomes routinely available.

Based on the slope $s$ of the $\log N - \log {\cal A}_X$ behaviour (or
equivalently on the benefit gained by stacking the signal from many
objects), we have found that a DM origin can be identified against an
astrophysical background in the case of decaying DM, but not for
annihilating DM. Though, in principle, combining the angular-dependence 
information and the stacked signal at different integration
angles gives clues as to the signal origin, the existing uncertainties
on the DM substructures (which may contribute significantly to the
annihilation signal) prevents us from reaching any clear conclusion.
Furthermore, whereas it has been sometimes argued that the angular
dependence can be used to disentangle DM decay from DM annihilation, 
we have shown for galaxy clusters that i) there is not strictly a
universal angular dependence for the annihilating signal, and ii) that
the very uncertainties on the DM substructures prevent us from predicting
exactly what this dependence is. Hence, decaying DM is the simplest
scenario to test and/or exclude, because it is the most sensitive to stacking
and the angular dependence of several objects can also be stacked (with
a rescaling).

The robustness of this analysis would benefit from a more thorough
investigation that takes into account various sources of  uncertainties,
such as the DM and gas profile modelling. For instance, the intrinsic
scatter observed in gas density profiles at the centre of clusters (due
to gas physics, interactions with the central galaxy and with the often
present active nucleus, the dynamical state of the cluster, etc.) could
affect the CR signal signature. Similarly a more systematic check
of the dependence of the DM annihilation signal on the clump parameters
(for the substructures' contribution) is certainly needed. The
calculations presented here have been done in the idealised scenario
where all clusters have a dominant signal component of the same nature,
whether it be DM decay, annihilation or astrophysical. The effect may be even
less clear if this is not the case. Nevertheless, this may be a useful
test that will complement those already suggested in the literature
(spectral feature, spatial-dependence of the signal, multi-wavelength
analysis, angular power spectrum). 

\begin{acknowledgements}
We thank J. Hinton and R. White for useful discussions. We thank
the referee for his/her careful reading and comments that helped clarify
the paper.
\end{acknowledgements}

\bibliographystyle{aa}
\bibliography{cluster_cr}
\end{document}